# Near-field Meta-optics

**Dongyoung Lee[1,†], Sehwan Kim[1,†], and Jisoo Kyoung[1,*]**

[1]Department of Physics, Dankook University, Cheonan 31116, Republic of Korea

[†]These authors contributed equally.

*kyoungjs@dankook.ac.kr

**Abstract:**

Metasurfaces have revolutionized compact wavefront control using planar, subwavelength structures. However, conventional meta-optical devices predominantly operate within a far-field paradigm, assuming electromagnetic decoupling between the source and metasurface, which limits control to post-emission wavefront shaping. Here, we define and experimentally demonstrate near-field meta-optics—a regime where strong source–structure coupling enables simultaneous control of emission and radiation. By integrating an inverse-designed dielectric metasurface directly within the near field of a terahertz photoconductive antenna (PCA), we show that the metasurface co-defines the emission process itself. Our meta-PCA, incorporating a 50-μm-high metasurface—one-third the thickness required by reciprocal far-field designs—collapses emission from ~60° divergence to a sharp <10° forward beam, while enhancing on-axis intensity 50-fold compared to bare GaAs. Unlike far-field metasurfaces that typically trade efficiency for thinness while remaining laterally large, our device achieves extreme compactness in both dimensions. Remarkably, it exceeds the outcoupling efficiency of a bulky, millimeter-scale silicon lens by 10%, despite a volume reduction of over three orders of magnitude. These results establish near-field meta-optics as a transformative paradigm for developing high-efficiency, ultra-compact on-chip photonic systems across the electromagnetic spectrum.

Meta-optics has transformed photonics by replacing bulky refractive components with planar, subwavelength-engineered surfaces.[1–8] Yet despite its rapid progress, the field has been built upon a largely implicit assumption: the optical source and the metasurface are electromagnetically decoupled. In nearly all established implementations, the metasurface resides in the far field of the emitter, where the incident wavefront is predefined and unaffected by the structure itself (Fig. 1(a)). Within this doctrine, meta-atoms function as locally independent phase shifters[9–12], and reciprocity provides the central design principle—beam shaping is achieved by engineering the reverse focusing problem.[13–19] While this far-field doctrine has catalyzed remarkable progress, it simultaneously constrains the operational boundaries of conventional meta-optics.

When a metasurface is brought into the electromagnetic near-field of an emitter, the foundational assumptions of the far-field paradigm no longer apply. The incident field is no longer fixed. The metasurface is no longer a passive wavefront modifier. Instead, source and structure become inseparably coupled: altering the metasurface modifies the emission process itself (Fig. 1(b)). The long-standing conceptual separation between emitter and optical component collapses. We term this regime near-field meta-optics.

In near-field meta-optics, radiation is engineered during its formation rather than after its release into free space. The metasurface participates directly in defining the emitter's radiation characteristics by reshaping the local electromagnetic environment. As a result, metasurface design can no longer rely on reciprocal phase mapping or periodic unit-cell approximations. The emitter and metastructure must instead be treated as a single, self-consistent system, rendering full-system inverse design essential rather than optional.

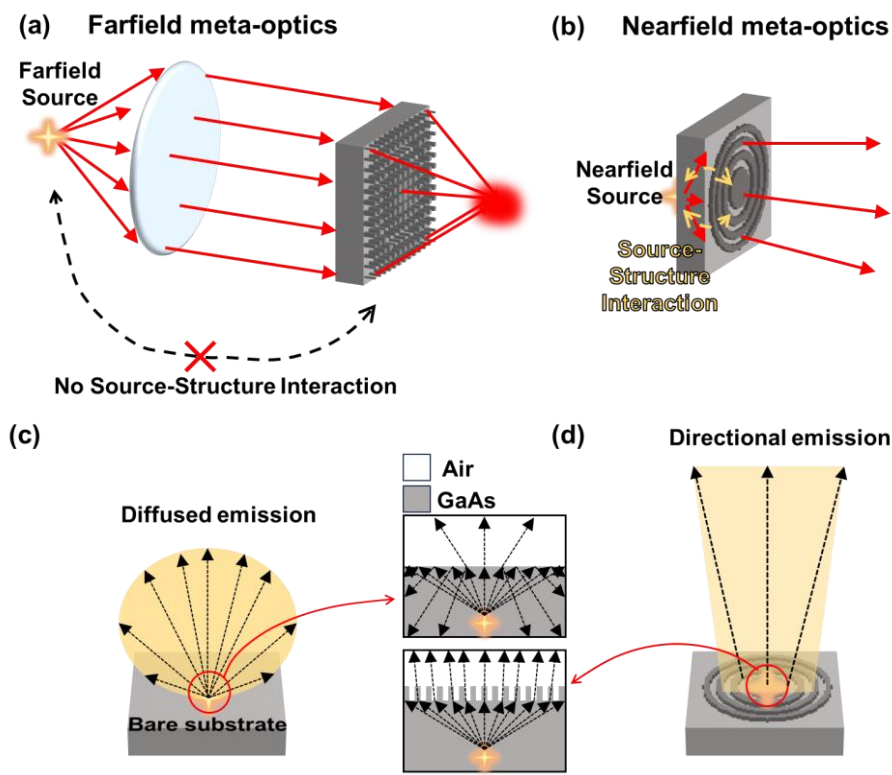

Figure 1. (a) Schematic illustration of far-field meta-optics, where the metasurface operates independently of the source (emitter). (b) Schematic illustration of near-field meta-optics, in which the source (emitter) and the metasurface form a strongly coupled system. (c) Diffusive terahertz emission and severe total internal reflection caused by the large refractive-index mismatch at the GaAs–air interface in a conventional photoconductive antenna. (d) Highly directive emission and suppression of total internal reflection enabled by the near-field monolithic metasurface integrated on GaAs.

This regime is particularly consequential for emitters embedded in high-index media, where extreme divergence and total internal reflection severely limit outcoupling (Fig. 1(c)). In such systems, reciprocal far-field strategies would demand physically inaccessible numerical apertures and impractically large apertures. Near-field meta-optics offers a fundamentally different pathway: by exploiting strong source–structure coupling, it reshapes radiation at its origin, reduces the required device footprint, and unlocks degrees of freedom inaccessible to conventional metasurface architectures.

Here we experimentally establish near-field meta-optics using a terahertz photoconductive antenna integrated with a monolithic, inverse-designed dielectric metasurface fabricated directly on the opposite side of a GaAs substrate (Fig. 1(d)). Operating within a few wavelengths of the dipolar emission region, the device transforms highly divergent radiation into a narrow forward-directed beam and significantly enhances forward emission relative to a bare substrate. The total outcoupling efficiency reaches that of a conventional bulky silicon lens, despite the metasurface being only one-third the height required in far-field implementations and replacing a centimeter-scale lens with a millimeter-scale monolithic platform. By demonstrating that metasurfaces can co-define, rather than merely modify, optical emission, this work establishes near-field meta-optics as a distinct design philosophy that transcends the limitations of far-field meta-optics and enables ultra-compact, high-efficiency photonic systems.

## Results and Discussion

**Inverse Design of THz Near-field Metasurface**

A conventional terahertz photoconductive antenna (PCA) consists of metallic electrodes patterned on a semi-insulating GaAs substrate.[20–22] When a femtosecond laser pulse illuminates the electrode gap under an applied bias, a transient photocurrent is generated directly beneath the GaAs surface, giving rise to broadband terahertz radiation. However, GaAs exhibits a very high refractive index of approximately 3.4 in the terahertz regime. As a consequence, the majority of the generated radiation undergoes total internal reflection at the GaAs–air interface and fails to escape the substrate, while the outcoupled portion exhibits an extremely large divergence angle (Fig. 1(c)).

To mitigate this limitation, thick silicon lenses are commonly attached to the backside of the GaAs substrate.[23] Silicon has a refractive index closely matched to that of GaAs and can be readily fabricated into a hyper-hemispherical geometry. When placed in direct contact with the substrate, the silicon lens suppresses internal reflection, enables efficient coupling of terahertz radiation into the far field, and simultaneously reduces beam divergence. Despite their effectiveness, such silicon lenses are inherently bulky, with radii on the order of several tens of millimeters. Moreover, their high refractive index still leads to substantial internal reflections, and their performance critically depends on precise mechanical alignment between the emission region and the lens center.

Here, we pursue a fundamentally different approach. Instead of attaching an external bulk optic, we directly pattern a dielectric monolithic metasurface onto the backside of the GaAs substrate opposite the photoconductive electrodes. This strategy eliminates alignment requirements altogether and enables a lens-integrated photoconductive antenna—hereafter referred to as a meta-PCA—that surpasses the efficiency of conventional silicon-lens-based systems while operating entirely within the near-field meta-optical regime (Fig. 1(d)).

To identify an optimal metasurface design, we performed full-wave finite-difference time-domain (FDTD, Lumerical, Ansys). simulations. Given the substrate thickness of approximately 600 μm, the metasurface resides well within the electromagnetic near-field of the terahertz emitter. As a result, the presence of the metasurface directly modifies the dipolar radiation process, rather than merely shaping a predefined incident wavefront.

Because the metasurface participates in the emission process itself, the incident phase distribution across the metasurface aperture is neither fixed nor smoothly varying. Instead, it is strongly correlated with the local geometry of the metastructure, rendering conventional unit-cell-based or locally periodic design strategies invalid. For this reason, we adopted a global inverse-design approach[24–26] in which the dipole source and the entire metasurface were treated as a single, self-consistent system. Specifically, a binary freeform dielectric metasurface was optimized using a genetic algorithm[27–32] directly coupled to FDTD simulations.

The transient photocurrent responsible for terahertz generation was modeled as a horizontal electric dipole located 1 μm beneath the GaAs surface. Owing to the symmetry of the horizontal dipole, anti-symmetric and symmetric boundary conditions were applied along the x- (perpendicular to the dipole orientation) and y- (parallel to the dipole orientation) directions, respectively, allowing the simulation domain to be reduced to a single quadrant. To further accelerate convergence of the inverse design, we additionally imposed near-circular symmetry on the metasurface geometry.

The objective of the inverse design was to efficiently extract terahertz radiation generated inside the high-index GaAs substrate into free space with minimal divergence. Accordingly, the figure of merit (FOM) was defined as the maximization of the far-field intensity of the zeroth-order transmission from GaAs substrate, corresponding to forward-directed radiation. The target operating frequency was set to 1 THz. Considering that the wavelength of 1-THz radiation inside GaAs is about 88 μm, and taking into account the constraints of the GaAs deep-etching fabrication process, the unit cell size of the binary metasurface was chosen to be 20 μm. Because the dipole emitter and the metasurface are separated by only a few wavelengths, a metasurface diameter of approximately 5 mm was sufficient to capture and reshape the relevant near-field radiation. The inverse design was carried out using a genetic algorithm with a population size of 60 individuals per generation and a total of 267 generations.

The height of the meta-atoms is a critical parameter that strongly influences metasurface performance. In conventional far-field meta-optics, where a full 0–2π phase coverage is required, meta-atoms with a refractive index of 3.4 typically require a minimum height of approximately 150 μm at 1 THz.[19,33] Although increasing the height generally reduces transmission due to propagation loss, such large aspect ratios are unavoidable to achieve the required phase modulation. In contrast, we performed inverse design optimizations under identical conditions for near-field metasurfaces with heights of 25, 50, 75, 100, and 150 μm. Among these, the height of 50 μm yielded the best performance. Notably, the optimal height of 50 μm is nearly one-third of that typically required for far-field metasurfaces, highlighting the fundamentally different operating principles of near-field meta-optics.

Unlike far-field metasurfaces that rely on complete 0–2π phase accumulation, near-field meta-optics operates through localized coupling between the emitter and the metasurface within a laterally confined region before forming the far-field beam. As a result, beam shaping is achieved primarily through enhanced radiation extraction and angular redistribution rather than phase-only engineering. This distinction allows efficient operation with significantly reduced meta-atom height, as excessive thickness introduces propagation loss and scattering without providing additional benefit in the near-field regime. This reduced structural height also offers substantial advantages from a fabrication standpoint compared with conventional far-field metasurfaces.

**Performance of Optimized Near-field Metasurface**

Figure 2(a) shows the optimized metasurface with a height of 50 μm, together with the evolution of FOM, normalized to its maximum value, over successive generations. Owing to the imposed near-circular symmetry, optimization exhibits rapid and stable convergence.

Figure 2(b) compares the angular radiation patterns of terahertz waves exiting the GaAs substrate for an unpatterned (bare) case and for the case patterned with the optimized metasurface. The radiation is plotted in the xz plane (left panel), which is parallel to the horizontal dipole orientation, and in the yz plane (right panel), which is perpendicular to it. The magnitude of the electric field is shown on a logarithmic scale ($-20\log|E|$), with red curves corresponding to the substrate patterned with the optimized metasurface and black curves to the bare substrate. In the absence of the metasurface, the large refractive-index mismatch between GaAs and air leads to severe internal reflection, resulting in a broad divergence of approximately 60° and weak outcoupling into free space. In contrast, the optimized metasurface dramatically suppresses divergence to below 10° while strongly enhancing forward-directed radiation. In particular, along the zeroth-order (forward) direction, the transmitted electric-field amplitude is enhanced by more than 27 dB relative to the bare substrate.

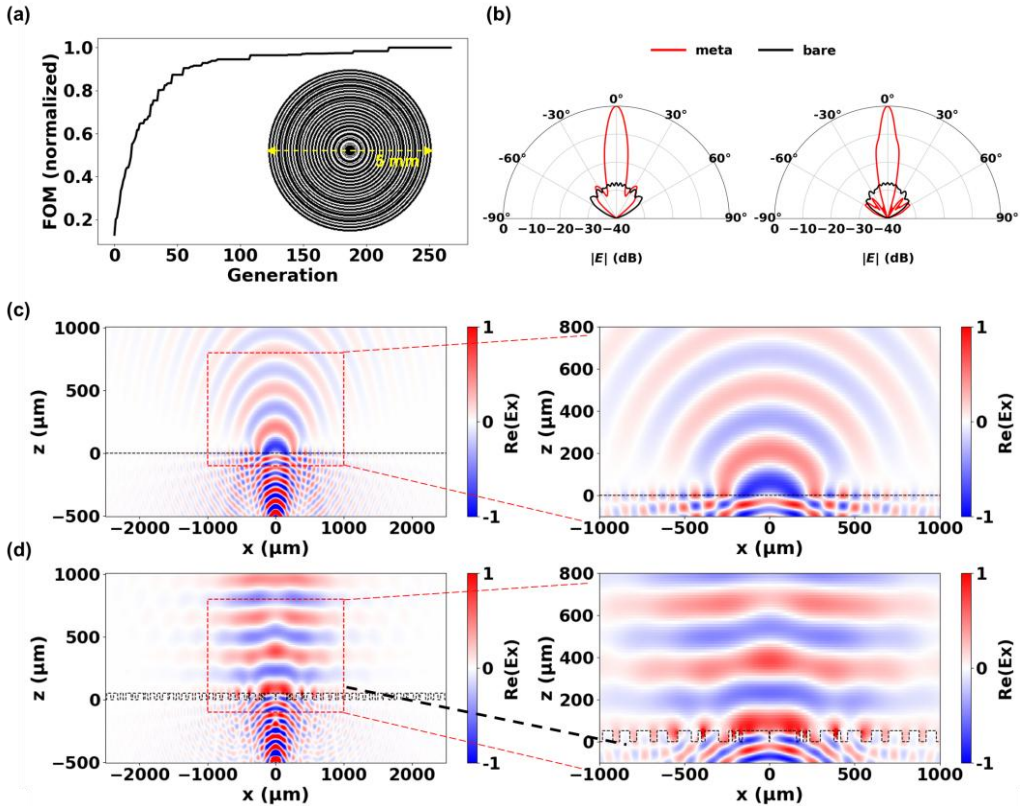

Figure 2. (a) Evolution of the normalized figure of merit (FOM) during inverse design optimization (inset: image of the optimized 5-mm-diameter near-field metasurface). (b) Simulated angular radiation patterns in the xz plane (left) and yz plane (right) for an unpatterned (bare) GaAs substrate and for a substrate patterned with the optimized

metasurface. (c) Near-field electric-field distribution for the bare configuration, showing strong total internal reflection and a highly divergent wavefront; the inset shows an enlarged view near the GaAs–air interface. (d) Near-field electric-field distribution for the optimized metasurface, exhibiting a nearly planar phase profile and suppressed total internal reflection; the inset shows an enlarged view near the interface.

To elucidate the phase-engineering capability of the near-field metasurface, Fig. 2(c) and Fig. 2(d) show the electric-field distributions in the near-field regime for the bare substrate and for the substrate patterned with the optimized metasurface, respectively. In the absence of a metasurface (bare substrate case) (Fig. 2(c)), radiation generated by the dipole inside the GaAs substrate undergoes strong reflection at the GaAs–air interface, and the transmitted wavefront exhibits pronounced spherical curvature, consistent with large angular divergence. By contrast, in the presence of the metasurface (Fig. 2(d)), interface reflections are strongly suppressed, and the emerging wavefront displays an almost planar phase profile immediately after passing through the metasurface. These results directly demonstrate that the near-field metasurface simultaneously reduces reflection and flattens the phase front with reshaping the radiation process itself rather than merely modifying an already formed beam.

**Fabrication and Experimental Characterization of Near-field Metasurface**

To implement the inverse-designed metasurface on an actual photoconductive antenna (PCA), we carried out a double-sided fabrication process on a GaAs substrate. The fabricated meta-PCAs were characterized using a home-built terahertz time-domain spectroscopy (THz-TDS) system. To minimize alignment-related ambiguities, we adopted a collinear emitter–detector geometry rather than a conventional THz-TDS configuration based on off-axis parabolic mirrors. Details of the laser system and optical layout are provided in the Ref.[34].

As the detector, we employed a commercial photoconductive antenna (PCA-40-05-10-800-h, BATOP) combined with a TPX(polymethylpentene) lens (CTL-D25mm, focal length 42 mm, BATOP). To enable a direct and unbiased comparison among a conventional silicon-lens-coupled emitter (a GaAs based PCA coupled to a bulk silicon lens attached to the substrate backside), a bare PCA (a GaAs based PCA without any silicon lens or metasurface), and the meta PCA, the detector assembly was kept fixed while only the emitter was exchanged.

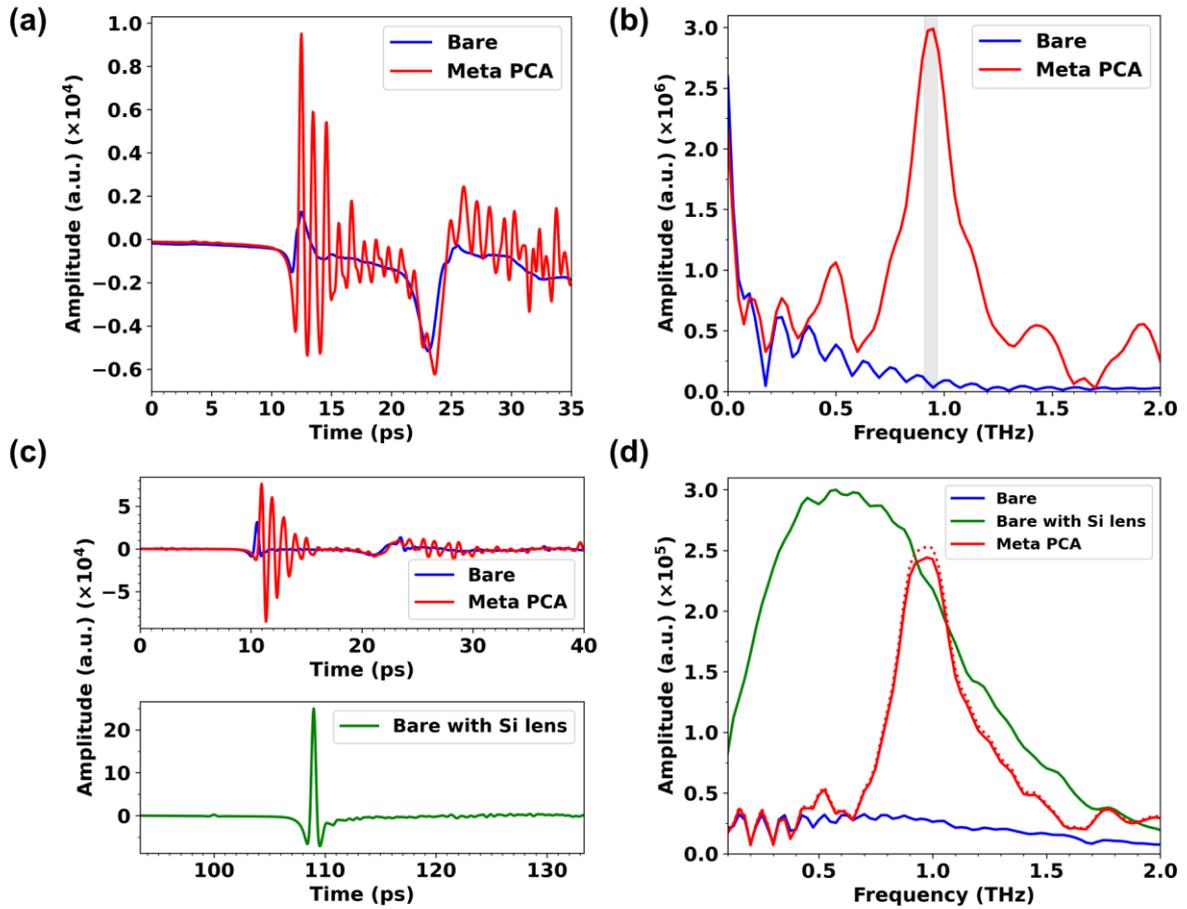

Figure 3. (a) Zeroth-order terahertz emission measured from the bare PCA and the meta PCA without additional collimating optics. (b) Frequency-domain spectra obtained by Fourier transforming the time-domain signals in (a); the shaded region highlights the spectral window used for quantitative comparison near the meta PCA resonance. (c) Time-domain terahertz signals measured after collimation: (top) bare PCA and meta PCA using a TPX lens (f = 32.5 mm), and (bottom) a conventional silicon-lens-coupled PCA emitter. (d) Corresponding frequency-domain spectra obtained by Fourier transforming the signals in (c). The meta PCA exhibits a stronger peak near 1 THz than the conventional bulk silicon lens. The dashed curve indicates the meta PCA spectrum corrected for Fresnel reflection losses at the TPX lens surfaces.

Figure 3(a) presents the time-domain terahertz signals measured at a distance of 110 mm from the emitter without any additional collimating optics. The detector system is optimized to collect near-collimated radiation, and given the sufficiently large emitter–detector separation, the measured signal can be regarded as dominated by the zeroth-order radiation beam. The red and blue traces correspond to the meta PCA and bare PCA, respectively. While the bare PCA produces only a weak detected signal, the meta PCA generates a markedly stronger quasi-continuous-wave terahertz response.

The corresponding frequency-domain spectra, obtained via Fourier transformation of the time-domain signals, are shown in Fig. 3(b). The meta PCA exhibits a pronounced spectral peak centered at 0.95 THz, in excellent agreement with the design target frequency of 1 THz, demonstrating close correspondence between numerical optimization, fabrication, and experiment. In contrast, the bare PCA displays no distinct resonance and follows a typical 1/f-like spectral behavior with enhanced low-frequency components.

To quantitatively evaluate the enhancement of forward emission, we averaged the spectral amplitudes over the shaded frequency window (0.91–0.97 THz) indicated in Fig. 3(b). Within this band, the meta PCA exhibits an approximately 57-fold increase in forward-emitted electric-field amplitude relative to the bare PCA.

We next benchmark the performance of the meta PCA against a conventional terahertz photoconductive antenna coupled to a bulk silicon lens. The silicon lens employed in this study has a diameter of 20 mm and is designed to collimate the highly divergent radiation emitted from a bare PCA. As a result, the terahertz beam becomes approximately collimated immediately after passing through the silicon lens, enabling direct detection without additional optical components. Because efficient coupling critically depends on precise positioning, the silicon lens was mounted on a three-axis translation stage and carefully aligned by maximizing the detected terahertz main peak. The resulting time-domain signal is shown in the lower panel of Fig. 3(c).

In contrast, radiation emitted from both the bare PCA and the meta PCA remains divergent. For these cases, a TPX lens (focal length 32.5 mm, LTA-D25.4-F32.5, BATOP) was inserted between the emitter and detector to collimate the beam prior to detection. Because the divergence angles of the two emitters differ substantially, a z-scan was performed by translating the TPX lens along the optical axis to identify the position that maximized the detected time-domain peak for each emitter. Owing to its significantly reduced divergence, the optimal TPX lens position for the meta PCA was found to be 1.12 mm closer to the emitter than for the bare PCA.

The corresponding time-domain signals measured under these optimized conditions are shown in the upper panel of Fig. 3(c). Consistent with the measurements without the TPX lens, the meta PCA exhibits a strong quasi-continuous-wave oscillation, whereas the bare PCA produces only a weak signal. To facilitate spectral comparison, the time-domain signals in Fig. 3(c) were Fourier transformed, and the resulting frequency-domain spectra are presented in Fig. 3(d). The conventional silicon-lens-coupled emitter (green curve) shows a broadband response with a peak near 0.6 THz, as commonly observed in silicon-lens-assisted terahertz PCAs. In contrast, when the silicon lens is removed (bare PCA case) and only the TPX lens is used (blue curve), the detected signal drops to approximately 8% of that obtained with the silicon lens. This drastic reduction reflects the severe outcoupling loss at the GaAs–air interface and the inability of the TPX lens to collect the strongly divergent radiation emitted from the bare PCA.

The emission spectrum of the meta PCA is shown as the red curve in Fig. 3(d). A pronounced and narrow peak is observed near the target frequency of 1 THz as anticipated. Remarkably, the detected signal at this peak exceeds that obtained using the conventional bulk silicon lens. Quantitatively, the peak amplitude reaches approximately 110% of that measured from the silicon-lens-coupled PCA. It is important to emphasize that this enhancement is achieved despite a reduction in the intrinsic dipole power induced by the optimized near-field metasurface. In other words, the total emitted power of the dipole itself is suppressed by the strong source–metasurface coupling, yet the collected far-field signal increases by more than 10% compared to the bulk silicon lens. This directly indicates that the near-field metasurface effectively extracts and redirects a substantially larger fraction of the generated terahertz radiation into the forward-propagating mode.

In addition, unlike the silicon-lens configuration, the meta PCA measurements require an additional TPX lens for beam collimation prior to detection. This introduces unavoidable Fresnel reflection losses at the TPX lens surfaces. Assuming normal incidence and a refractive index of 1.47 for TPX, the transmission amplitude through the lens is approximately 96.4%. Accounting for this loss, the terahertz amplitude incident on the detector would be expected to be roughly 3.6% higher than the measured value. The corresponding reflection-corrected estimate is indicated by the dashed curve in Fig. 3(d).

## Conclusion

In conclusion, this work establishes near-field meta-optics as a fundamentally distinct regime of optical design, governed by principles that differ from those of conventional far-field meta-optics. Whereas far-field metasurface offer reduced thickness, high numerical aperture, and multifunctionality at the cost of efficiency, we show that operating metasurfaces within the electromagnetic near field of the source can overcome this long-standing trade-off. By exploiting strong source–metasurface coupling, our platform simultaneously compresses both the vertical and lateral dimensions while enhancing radiation extraction beyond what is achievable with conventional bulk optics. Moreover, unlike silicon lenses that require precise three-dimensional alignment behind high-index emitters, monolithic integration directly onto the GaAs substrate eliminates alignment constraints. Although demonstrated here in the terahertz regime, our approach can be extended to cases where the entire device stack is on the order of the wavelength in thickness and the emitter is embedded within a high-refractive-index medium, such as OLEDs, PeLEDs, and TMD monolayer LEDs.[35,36] Near-field meta-optics thus provides a new design framework for ultra-compact, high-efficiency photonic systems.

**References**

1. Yu, N. *et al.* Light propagation with phase discontinuities: generalized laws of reflection and refraction. *Science (1979).* **334**, 333–337 (2011).
2. Khorasaninejad, M. & Capasso, F. Metalenses: Versatile multifunctional photonic components. *Science (1979).* **358**, (2017).
3. Chen, W. T., Zhu, A. Y., Sisler, J., Bharwani, Z. & Capasso, F. A broadband achromatic polarization-insensitive metalens consisting of anisotropic nanostructures. *Nat. Commun.* **10**, 355 (2019).
4. Joo, W.-J. *et al.* Metasurface-driven OLED displays beyond 10,000 pixels per inch. *Science (1979).* **370**, 459–463 (2020).
5. Pan, M. *et al.* Dielectric metalens for miniaturized imaging systems: progress and challenges. *Light Sci. Appl.* **11**, 195 (2022).
6. Schulz, S. A. *et al.* Roadmap on photonic metasurfaces. *Appl. Phys. Lett.* **124**, (2024).
7. Kuznetsov, A. I. *et al.* Roadmap for Optical Metasurfaces. *ACS Photonics* **11**, 816–865 (2024).
8. Gopakumar, M. *et al.* Full-colour 3D holographic augmented-reality displays with metasurface waveguides. *Nature* **629**, 791–797 (2024).
9. Yu, N. & Capasso, F. Flat optics with designer metasurfaces. *Nat. Mater.* **13**, 139–150 (2014).
10. Khorasaninejad, M. *et al.* Metalenses at visible wavelengths: Diffraction-limited focusing and subwavelength resolution imaging. *Science (1979).* **352**, 1190–1194 (2016).
11. Kildishev, A. V, Boltasseva, A. & Shalaev, V. M. Planar Photonics with Metasurfaces. *Science (1979).* **339**, 1232009 (2013).
12. Arbabi, A., Horie, Y., Ball, A. J., Bagheri, M. & Faraon, A. Subwavelength-thick Lenses with High Numerical Apertures and Large Efficiency Based on High Contrast Transmitarrays. *Nat. Commun.* **6**, 1–10 (2015).
13. Kim, Y.-B., Cho, J.-W., Bae, D. & Kim, S.-K. Single-unit metalens integrated micro light-emitting diodes. *Current Applied Physics* **67**, 85–92 (2024).
14. Arbabi, A., Briggs, R. M., Horie, Y., Bagheri, M. & Faraon, A. Efficient dielectric metasurface collimating lenses for mid-infrared quantum cascade lasers. *Opt. Express* **23**, 33310–33317 (2015).
15. Chen, E. *et al.* Broadband beam collimation metasurface for full-color micro-LED displays. *Opt. Express* **32**, 10252–10264 (2024).
16. Suzuki, T., Endo, K. & Kondoh, S. Terahertz metasurface ultra-thin collimator for power enhancement. *Opt. Express* **28**, 22165–22178 (2020).

17. Yu, Q. *et al.* All-Dielectric Meta-lens Designed for Photoconductive Terahertz Antennas. *IEEE Photonics J.* **9**, 1–9 (2017).
18. Roy, T. R., Sunder Meetei, T. & Yu, N. E. Design of a metalens for beam collimation and angular amplification in optical phased array devices. *Opt. Express* **32**, 34344–34355 (2024).
19. Yu, Q. *et al.* All-Dielectric Meta-lens Designed for Photoconductive Terahertz Antennas. *IEEE Photonics J.* **9**, 1–9 (2017).
20. Neu, J. & Schmuttenmaer, C. A. Tutorial: An introduction to terahertz time domain spectroscopy (THz-TDS). *J. Appl. Phys.* **124**, (2018).
21. Koch, M., Mittleman, D. M., Ornik, J. & Castro-Camus, E. Terahertz time-domain spectroscopy. *Nature Reviews Methods Primers* **3**, 48 (2023).
22. Burford, N. M. & El-Shenawee, M. O. Review of terahertz photoconductive antenna technology. *Optical Engineering* **56**, 010901 (2017).
23. Van Rudd, J. & Mittleman, D. M. Influence of substrate-lens design in terahertz time-domain spectroscopy. *Journal of the Optical Society of America B* **19**, 319 (2002).
24. Li, Z., Pestourie, R., Lin, Z., Johnson, S. G. & Capasso, F. Empowering Metasurfaces with Inverse Design: Principles and Applications. *ACS Photonics* **9**, 2178–2192 (2022).
25. Elsawy, M. M. R., Lanteri, S., Duvigneau, R., Fan, J. A. & Genevet, P. Numerical Optimization Methods for Metasurfaces. *Laser Photon. Rev.* **14**, 1900445 (2020).
26. Jeong, J.-Y., Latif, S. & So, S. A Tutorial on Inverse Design Methods for Metasurfaces. *Current Optics and Photonics* **8**, 531–544 (2024).
27. Zhang, J., Wang, G., Wang, T. & Li, F. Genetic Algorithms to Automate the Design of Metasurfaces for Absorption Bandwidth Broadening. *ACS Appl. Mater. Interfaces* **13**, 7792–7800 (2021).
28. Jafar-Zanjani, S., Inampudi, S. & Mosallaei, H. Adaptive Genetic Algorithm for Optical Metasurfaces Design. *Sci. Rep.* **8**, 11040 (2018).
29. Wang, Y. *et al.* Genetic algorithm-enhanced design of ultra-broadband tunable terahertz metasurface absorber. *Opt. Laser Technol.* **170**, 110262 (2024).
30. Tao, E. *et al.* Neural network and genetic algorithm-driven inverse design of a full-phase metasurface for wavefront manipulation. *Journal of the Optical Society of America B* **42**, 2198–2206 (2025).
31. Katoch, S., Chauhan, S. S. & Kumar, V. A review on genetic algorithm: past, present, and future. *Multimed. Tools Appl.* **80**, 8091–8126 (2021).
32. Whiting, E. B., Kang, L., Jenkins, R. P., Campbell, S. D. & Werner, D. H. Broadband plasmonic chiral meta-mirrors. *Opt. Express* **31**, 22415–22423 (2023).
33. Seo, D.-J. & Kyoung, J. Shape dependence of all-dielectric terahertz metasurface. *Opt. Express* **30**, 38564 (2022).
34. Yu-Jin Nam & Jisoo Kyoung. Software-based Simple Lock-in Amplifier and Built-in Sound Card for Compact and Cost-effective Terahertz Time-domain Spectroscopy System. *Current Optics and Photonics* **7**, 683–691 (2023).
35. Tan, Z.-K. *et al.* Bright light-emitting diodes based on organometal halide perovskite. *Nat. Nanotechnol.* **9**, 687–692 (2014).
36. Withers, F. *et al.* Light-emitting diodes by band-structure engineering in van der Waals heterostructures. *Nat. Mater.* **14**, 301–306 (2015).